\documentclass[twocolumn,showpacs,preprintnumbers,amsmath,amssymb]{revtex4}

\usepackage{graphicx}
\usepackage{dcolumn}
\usepackage{bm}
\usepackage{color}

\usepackage{amssymb}
\usepackage{amsmath}
\usepackage{hyperref}

\newif\ifold             \oldtrue            
\def\be{\begin{eqnarray}}
\def\ee{\end{eqnarray}}

\begin{document}

\title{The Classical and Commutative Limits of noncommutative Quantum
Mechanics:\\ A Superstar $\bigstar$ Wigner-Moyal Equation}

\author{Ardeshir Eftekharzadeh and B. L. Hu}
\email [Electronic addresses]{: eftekhar@wam.umd.edu,
hub@physics.umd.edu}

\affiliation{Department of Physics, University of Maryland,\\
College Park, Maryland 20742-4111}
\date{May 5, 2005\\
{\it An Invited paper at the DICE04 meeting, Piombino, Italy.
September 04. To appear in Brazilian J. Physics}}

\begin{abstract} We are interested in the similarities and differences
between the quantum-classical (Q-C) and the
noncommutative-commutative (NC-Com) correspondences. As one useful
platform to address this issue we derive the superstar
Wigner-Moyal equation for noncommutative quantum mechanics (NCQM).
A superstar $\bigstar$-product combines the usual phase space
$\ast$ star and the noncommutative $\star$ star-product. Having
dealt with subtleties of ordering present in this problem we show
that the Weyl correspondence of the NC Hamiltonian has the same
form as the original Hamiltonian, but with a non-commutativity
parameter $\theta$-dependent, momentum-dependent shift in the
coordinates. Using it to examine the classical and the
commutative limits, we find that there exist qualitative
differences between these two limits. Specifically, if $\theta
\neq 0$ there is no classical limit. Classical limit exists only
if $\theta \rightarrow 0$ at least as fast as $\hbar \rightarrow
0$, but this limit does not yield Newtonian mechanics, unless the
limit of $\theta/\hbar$ vanishes as $\theta \rightarrow 0$. For
another angle towards this issue we formulate the NC version of
the continuity equation both from an explicit expansion in orders
of $\theta$ and from a Noether's theorem conserved current
argument. We also examine the Ehrenfest theorem in the NCQM
context.
\end{abstract}

\maketitle

\noindent{\bf Aim} In this program of investigation we ask the
question whether there is any structural similarity or conceptual
connection between the quantum-classical (Q-C) and the
noncommutative-commutative (NC-Com) correspondences.  We want to
see if our understanding  of the quantum-classical correspondence
acquired in the last decade can aid us in any way to understand
the {\it physical} attributes and meanings of a noncommutative
space from the vantage point of the ordinary commutative space. We
find that the case of quantum to classical transition in the
context of noncommutative geometry is quite different from that in
the ordinary (commutative) space. Specifically, if $\theta \neq
0$ there is no classical limit. Classical limit exists only if
$\theta \rightarrow 0$ at least as fast as $\hbar \rightarrow 0$,
but this limit does not yield Newtonian mechanics, unless the
limit of $\theta/\hbar$ vanishes as $\theta \rightarrow 0$. We
make explicit this relationship by deriving a superstar $\bigstar$
Wigner-Moyal equation for noncommutative quantum mechanics (NCQM)
and identifying the difference between the classical and the
commutative limits. A superstar $\bigstar$-product combines the
usual phase space $\ast$ star and the noncommutative
$\star$-product \cite{starprod}.

In this paper we focus on the nature of the commutative and
classical limits of noncommutative quantum physics.  We point out
some subtleties which arise due to the ordering problem. When
these issues are properly addressed we show that the classical
correspondent to the NC Hamiltonian is indeed one with a
$\theta$-dependent, momentum-dependent shift in the coordinates.
For another angle towards this issue we formulate the NC version
of the continuity equation both from an explicit expansion in
orders of $\theta$ and from a Noether's theorem conserved current
argument. We also examine the Ehrenfest theorem in the NCQM
context.

\section{Criteria for Classicality}

We open this discussion by examining the quantum to classical
(Q-C) transition issue which is probably more familiar to us than
the noncommutative to commutative (NC-Com) transition. We begin by
listing the criteria related to the Q-C issue so that we can see
the possibilities in how to approach the NC-Com issue, if there
is some analogy we can draw. In fact the focus of this paper is
to ask if any such analogy or parallel exists, both conceptually
and structurally. (The following is excerpted from \cite{Drexel})

A quick sampling of discussions in quantum mechanics and
statistical mechanics textbooks reveals a variety of seemingly
simple and straightforward criteria and  conditions
for classicality. For example, one can loosely associate:\\
1) $\hbar \rightarrow 0$\\
2) WKB approximation, which ``gives the semiclassical limit"\\
3) Ehrenfest Theorem, ``quantum expectation follows a classical equation of motion" \\
3) Wigner function, ``behaves like a classical distribution function"\\
4) high temperature limit: ``thermal=classical"\\
5) Uncertainty Principle: a system ``becomes classical" when this
is no longer obeyed\\
6) coherent states:  the `closest' to the classical\\
7) systems with large quantum number $n \rightarrow \infty$
(correspondence principle)\\
8) systems with large number of components $1/N \rightarrow 0$.\\

Each of these conditions contains only some partial truth and
when taken on face value can be very misleading. Many of these
criteria hold only under special conditions. They can
approximately define the classical limit only when taken together
in specific ways. To understand the meaning of classicality it is
important to examine the exact meaning of these criteria,  the
conditions of their validity and how they are related to each
other.

We can divide the above conditions into four groups, according to
the different issues behind these criteria:\\
a) quantum interference,\\
b) quantum and thermal fluctuations,\\
c) choice of special quantum states,\\
d) meaning of the large $n$ and $N$ limits.

The first two groups of issues were discussed in \cite{Drexel}
using the paradigm of quantum open systems. The first set of
issues a) was discussed in the context of quantum cosmology by
Habib and Laflamme \cite{HabLaf}. They asserted that decoherence
is needed for the WKB Wigner function to show a peak in phase
space indicating the correlation between the physical variables
and their canonical conjugates which defines a classical
trajectory. This clarifies the loose connection of WKB, Wigner
function and classicality. For issue b), for ordinary systems the
time for thermal fluctuations to overtake quantum fluctuations is
also related to the time of decoherence. But a decohered system is
not necessarily classical. There is a quantum statistical regime
in between. This set of issues was addressed by Hu and Zhang
\cite{HuZha}. (See also \cite{AnaHal,ZPH}.) They derived an
uncertainty principle for a quantum open system at finite
temperature which interpolates between the (zero temperature)
quantum Heisenberg relation and the high temperature result of
classical statistical mechanics. This was useful for clarifying
the sometimes vague notions of quantum, thermal and classical.

In our current investigation we wish to use what was learned in
the last decade in Q-C to inquire about a simple yet important
issue, namely, under what conditions is the ordinary commutative
space a bona fide limit of NC space, or, what is the nature of
the NC-Com transition?

Recall for QM:
\begin{eqnarray}
 [\hat{x}^i, \hat{p}_j] = i \hbar \delta^i_j
 \end{eqnarray}
whereas for NC geometry, two spatial coordinates $x^i, x^j$
satisfy the relation
\begin{eqnarray}
[\hat{x}^i, \hat{x}^j]=i\theta^{ij} \label{nc}
\end{eqnarray}
We will refer to $\theta^{ij}$ or simply $\theta$ as the
non-commutativity parameter.

{F}rom (\ref{nc}), we can see that the non-commutativity parameter
$\theta$ has the dimension of length squared. If the geometry of
space-time at a fundamental level is to be noncommutative then
one possible candidate for $\sqrt{\theta}$ is the Planck length.
This is how quantum gravity is linked with NCG, which also bears a
close relation to matrix models, quantum groups, M-theory and
string theory
\cite{CDS,konechny,seiberg,witten,douglas,nekrasov,Majid}.

Here we will actually work around the simplest criterion 1) $\hbar
\rightarrow 0$ limit in QM and ask the parallel question how the
$\theta \rightarrow 0$ limit would be different, and how these
two limits relate to each other. The place where both Q-C and
NC-Com share some nontrivial point of contact, at least formally,
is the Weyl correspondence between operators and c-functions, the
star product, the Wigner distribution, and the Wigner-Weyl
equation. This is the domain of semiclassical or semiquantal
physics. We will use this equation and the Wigner-Weyl
correspondence to explore the NC-Com and the Q-C transition.

\section{Quantum-Classical Correspondence}

The Wigner distribution function has found applications in kinetic
theory and has been instrumental in studying quantum coherence and
quantum to classical transitions. Star product arises from
considering the implications of Weyl transformation of quantum
canonical operators. (A good introduction to these topics can be
found in \cite{Liboff}. A succinct treatment of Moyal Bracket can
be found in an Appendix of \cite{Reichl}. Readers familiar with
these topics can skip to the next section.)

For simplicity, in what follows, we consider one dimensional
motion. The phase space canonical coordinates are denoted by $q$
and $p$ respectively for position and momentum dynamical variables
and their corresponding quantum mechanical operators are denoted
by $\hat{q}$ and $\hat{p}$.

\subsection{Weyl correspondence}


Weyl ~\cite{weyl} proposed that all dynamical variables be written
in terms of members of the Lie algebra of transformations given
by:
\begin{eqnarray}
\hat{U}(\lambda,\mu) = e^{i(\lambda \hat{q} + \mu \hat{p})/\hbar}
\end{eqnarray}
Let us define the set of \emph{phase-space operators} as the set
of all operators whose operator properties solely depends on
$\hat{q}$ and $\hat{p}$. Throughout this article, a member of this
set will be called a phase-space operator. One can show that for
such operators we can give the following representation:
\begin{eqnarray}
\hat{A}(\hat{q},\hat{p}) = \int d \lambda d \mu \;
\alpha(\lambda,\mu)e^{i(\lambda \hat{q} + \mu \hat{p})/\hbar}
\label{wc1}
\end{eqnarray}
$\alpha(\lambda,\mu)$ can be projected back to ($q,p$) space by
the inverse transformation:
\begin{eqnarray}
\alpha(\lambda,\mu) = \frac{1}{2\pi \hbar}\int dq \; dp \;
A_W(q,p)e^{-i(\lambda q + \mu p)/\hbar} \label{wc2}
\end{eqnarray}
where $A_W$ is called the Weyl correspondence of $\hat A$. We can
combine equations (\ref{wc1}) and (\ref{wc2}) to obtain:
\begin{eqnarray}
\nonumber \hat{A}(\hat{q},\hat{p}) = \frac{1}{(2\pi \hbar)^2}\int
dq dp \; d\lambda d \mu \; A_W(q,p) e^{i\left(\frac{\lambda
(\hat{q}-q) + \mu(\hat{p} -p)}{\hbar}\right)} \label{wc3}\\
\end{eqnarray}
The relationship defines a mapping from the set of functions of
phase-space variables to the set of phase-space operators. By
multiplying both sides of (\ref{wc3}), taking the trace of both
sides and making use of the fact that the $U$ transformations can
be inverted since
\begin{eqnarray}
\nonumber Tr[U(\lambda,\mu) U^{\dag}(\lambda',\mu')] = 2 \pi \hbar
\; \delta(\lambda - \lambda')\; \delta (\mu-\mu')
e^{i\frac{\lambda \mu - \lambda' \mu'}{2\hbar}}, \\ \label{uu}
\end{eqnarray}
we can find the inverse of the above mapping~\cite{harvey}. The
result is
\begin{eqnarray}
\nonumber A_W(q,p) = \frac{1}{2\pi \hbar}\int e^{i(\lambda q + \mu
p)/\hbar}\; Tr[U^{\dag}(\lambda,\mu)\hat{A}(\hat{q},\hat{p})]
\;d\lambda d\mu \\ \label{rwt}
\end{eqnarray}
In what follows we show that every phase-space operator denoted
by $\hat{A}(\hat{q},\hat{p})$ can be written as the mapping
represented by (\ref{wc3}). First we note that every such
operator is completely determined by its matrix elements taken
with respect to any complete basis. Let the set of position
eigenstates be such a basis. One can fully represent the operator
in question as $\langle x_1 | \hat{A} | x_2 \rangle$. Introduce
the following change of variables
\begin{equation}
x_1 = X + \Delta, \,\,\,  x_2 = X - \Delta
\end{equation}
with inverse
\begin{equation}
X = \frac{x_1 + x_2}{2}, \,\,\,  \Delta = \frac{x_1-x_2}{2}
\end{equation}
one can define
\begin{eqnarray}
\label{me}A(X,\Delta) \equiv \langle x_1 | \hat{A} | x_2 \rangle
\end{eqnarray}
Every operator can be written in such a way. Next, we use the
Fourier theorems to write:
\begin{eqnarray}
\langle x_1 | \hat{A} | x_2 \rangle = \frac{1}{2\pi\hbar}\int dp
A_W(X,p) e^{-i\Delta p/\hbar}
\end{eqnarray}
where the use of index $W$ and the connection with the Weyl
correspondence will be clarified shortly. Inserting an integral
over additional variables
\begin{eqnarray}
\langle x_1 | \hat{A} | x_2 \rangle &=& \frac{1}{2\pi\hbar}\int
dpdq \delta(X-q) A_W(q,p) e^{-i\Delta p/\hbar} \\ \nonumber &=&
\frac{1}{(2\pi\hbar)^2} \int dpdqd\lambda e^{i\lambda(X-q)/\hbar}
A_W(q,p) e^{-i\Delta p/\hbar}
\end{eqnarray}
we get
\begin{widetext}
\begin{eqnarray}
\langle x_1 | \hat{A} | x_2 \rangle  =\frac{1}{(2\pi\hbar)^2}
\int dpdqd\lambda A_W(q,p) e^{i\lambda\left(-q + \frac{x_1 +
x_2}{2}\right)/\hbar} e^{i(x_1-x_2)(\frac{\lambda}{2} - p)/\hbar}
\end{eqnarray}
\end{widetext}
Now insert another integral over \linebreak $\delta(x_1-x_2-\mu)$
to eliminate $(x_1-x_2)$. This latter Dirac delta function can be
written as  $\langle x_1 | x_2 + \mu\rangle$ which is equal to
$\langle x_1 | e^{i\mu \hat{p}/\hbar}| x_2 \rangle$. Once the
position eigen kets are inserted, one can write factors like
$\left(\; e^{i\lambda x_1/\hbar} \langle x_1| \;\right) $ as
$\left(\; \langle x_1 |e^{i\lambda \hat{q}/\hbar}\;\right) $.
Combining all of the above we have
\begin{widetext}
\begin{eqnarray}
\langle x_1 | \hat{A} | x_2 \rangle = \frac{1}{(2\pi\hbar)^2} \int
dp\, dq \,d\lambda\, d\mu A_W(q,p) e^{-i(\lambda q + \mu p)/\hbar}
 \langle x_1 | e^{\frac{i\lambda \hat{q}}{2\hbar}}
e^{i\mu \hat{p}/\hbar} e^{\frac{i\lambda \hat{q}}{2\hbar}} | x_2
\rangle
\end{eqnarray}
\end{widetext}
Now we know that the $|x_1 \rangle$ and $|x_2 \rangle$ were
arbitrary. If the operator properties of $\hat{A}$ solely depends
on $\hat{p}$ and $\hat{q}$, that is,  if the collection of all the
matrix elements of the type (\ref{me}) can fully describe the
operator $\hat{A}$, then the aforementioned state kets  can be
omitted from both sides of the equation. Then we can use the
Baker-Campbell-Hausdorff lemma to combine operators inside the
bra-ket into $e^{(\lambda \hat{q} + \mu \hat{p})/\hbar}$ and
therefore show that any phase-space operator can be written as
(\ref{wc3}). That is to say,  for every phase-space operator,
there is a function of the phase-space variables such that the
relationship (\ref{wc3}) holds. Thus the Weyl correspondence
represented by (\ref{wc3}) is an \emph{onto mapping} from the
space of functions into the space of phase-space operators.
Furthermore one can show that the Weyl correspondence is a
one-to-one mapping. To see that let us assume there are two
different functions, namely $A_W(q,p)$ and $A'_W(q,p)$ that map to
a single operator. That is
\begin{eqnarray}
&\,& \int dq dp d\lambda d \mu \; A_W(q,p) e^{i\left(\frac{\lambda
(\hat{q}-q) + \mu(\hat{p} -p)}{\hbar}\right)} \\ \nonumber &=&
\int dq dp \; d\lambda d \mu A'_W(q,p) e^{i\left(\frac{\lambda
(\hat{q}-q) + \mu(\hat{p} -p)}{\hbar}\right)}
\end{eqnarray}
Now one can use (\ref{uu}) to reverse both sides and by using the
properties of the Fourier transformation can show that $A_W(q,p)$
and $A'_W(q,p)$, are indeed identical. Therefore the Weyl
correspondence is a one-to-one and onto mapping from the set of
functions over the phase-space variables to the set of phase-space
operators as defined at the beginning of this subsection.

\subsection{Wigner Function}

Wigner distribution functions $W(q,p)$ in quantum systems are
meant to play the corresponding role of classical distributions in
classical kinetic theory. For a classical system in kinetic theory
and a positive-definite distribution function $P(q,p)$ of the
canonical variables $q,p$ in classical phase space,  we have
~\cite{Hillery:1983ms}:
\begin{eqnarray}
\langle A \rangle_{classical} = \int A(q,p) P(q,p) dq dp
\end{eqnarray}
Let us assume that the quantum system is described by the wave
function $\psi(x) = \langle x |\psi \rangle$. One can define
\begin{eqnarray}
\label{wc4}A(q,p) &=& 2 \int dx \;e^{2ixp/\hbar} \;\langle q - x |
\hat{A}(\hat{q},\hat{p}) | q + x \rangle \\
W(q,p) &=& \frac{1}{\pi\hbar} \int dx\; e^{-2ixp/\hbar}\; \psi^{*}
(q-x) \psi(q+x)
\end{eqnarray}
where $W(q,p)$ is called the Wigner function and it can be shown
to have the following main property
\begin{eqnarray}
\langle \hat{A} \rangle_{quantum} = \int A(q,p) W(q,p) dq dp
\end{eqnarray}
Note here that $\hat{A}$ is a phase-space operator. To show that
the transformation defined by Eq.~(\ref{wc4}) is equivalent to the
Weyl correspondence we use Eq.~(\ref{rwt}) to obtain
 \begin{eqnarray}
 A_W(q,p) =\frac{1}{2\pi \hbar} \int d\lambda d\mu \;e^{i(\lambda q
 + \mu p)/\hbar}\; Tr[U^{\dag}(\lambda,\mu) \hat{A}]. \label{wc5}
 \end{eqnarray}
 The factor involving the trace can be rewritten as,
 \begin{eqnarray}
 \nonumber Tr[U^{\dag}(\lambda,\mu) \hat{A}] &=& \int dq' \langle
 q' | e^{-i\lambda \hat{q}/\hbar} e^{-i\mu \hat{p}/\hbar}
 e^{i\lambda \mu/2\hbar} A
 | q' \rangle \\ \nonumber
 \nonumber &=& \int dq' e^{-i\lambda q'/\hbar} e^{i\lambda
 \mu/2\hbar} \langle q' |
 e^{-i\mu \hat{p}/\hbar} A| q' \rangle \\\nonumber
 &=& \int dq' e^{-i\lambda q'/\hbar} e^{i\lambda \mu/2\hbar}
\langle q'- \mu | A | q' \rangle \\
 \end{eqnarray}
which can be substituted back in Eq. ~(\ref{wc5}) to obtain
 \begin{eqnarray}
&\;& \nonumber A_W(q,p) \\ \nonumber &=& \frac{1}{2\pi \hbar} \int
d\mu \; dq'\; e^{i \lambda
(q - q' +\mu/2)/\hbar} e^{i\mu p} \langle q' - \mu | A | q' \rangle  \\
\nonumber &=& \int d\mu \; dq'\;\delta(q'-q - \mu/2)  e^{i\mu p}
\langle q' -\mu | A | q' \rangle \\
 &=& \int d\mu \; e^{i\mu p} \langle q - \mu/2 | A
| q + \mu/2 \rangle .
\end{eqnarray}
With a slight change of variable to $x = \mu/2$ we have
\begin{eqnarray}
A_W(q,p) = 2 \int   e^{i2x p/\hbar} \langle q - x | A | q + x
\rangle dx,
\end{eqnarray}
which proves the equivalence of the two mappings.\\Note that the
above transformation has the following property
\begin{eqnarray}
\int dqdp \; A(q,p) = 2\pi \hbar Tr[\hat{A}(\hat{q},\hat{p})]
\end{eqnarray}
and that the Wigner distribution function is actually the Weyl
transformation of the density matrix operator $\hat{\rho} = |\psi
\rangle \langle \psi |$
\begin{eqnarray}
\nonumber \rho_W(q,p) &=& 2\pi \hbar W(q,p) \\
          &=& 2 \int e^{i 2\bar{q}p/c}
\langle q - \bar{q}| \psi \rangle \langle \psi | q +\bar{q}\rangle
d\bar{q}
\end{eqnarray}
Unlike the classical case, where a probabilistic interpretation of
the distribution function is possible, the Wigner function cannot
be interpreted as a probability distribution because in general it
is not everywhere positive. Let $| p \rangle$ and $| q \rangle$ be
the eigenkets of operators $\hat{p}$ and $\hat{q}$ with
eigenvalues $p$ and $q$ respectively and the system is in the
state denoted by $|\psi\rangle$ it can be easily shown that the
Wigner function has the following properties:
\begin{eqnarray}
\nonumber \int W(q,p) dq =\langle p |\psi\rangle\langle\psi| p
\rangle, \;\;\; \int W(q,p)dp = \langle
q|\psi\rangle\langle\psi|q\rangle \\
\end{eqnarray}
For completeness we note that for a mixed state the Wigner
function can be defined as
\begin{eqnarray}
W(q,p) &=& \frac{1}{\pi\hbar} \int dx\; e^{-2ixp/\hbar}\;
\rho(q-x,q+x)
\end{eqnarray}
\subsection{Phase space $\ast$-product and Wigner-Moyal equation}

Consider two dynamical variables $A$ and $B$ in  a classical
system. The statistical average of their product is obtained by
weighting it with the distribution function $P(q,p)$ given by
\begin{eqnarray}
\langle A B \rangle_{classical} = \int A(q,p) B(q,p) P(q,p) dqdp .
\end{eqnarray}
If $A$ and $B$ are quantum mechanical operators,  because of their
functional dependence on the non-commuting operators $\hat q$ and
$\hat p$
a different rule of multiplication, the star product, is needed.
The star product satisfies the following property \cite{starprod}
\begin{eqnarray}
\langle \hat{A}\hat{B} \rangle = \int A_W(q,p)* B_W(q,p) W(q,p)
dqdp
\end{eqnarray}
Alternatively,
\begin{eqnarray}
C(q,p) = (\hat{A}\hat{B})_W(q,p) = A_W(q,p)* B_W(q,p)
\end{eqnarray}
where the symbol $(\;. \;)_W$ for products of operators, stands
for the Weyl transformation of the enclosed operator inside. How
is the star product related to the ordinary algebraic product? To
find out we first use the Weyl analysis for the solution
\begin{eqnarray}
\nonumber C(q,p) =   \frac{1}{2\pi \hbar}\int e^{i(\lambda q + \mu
p)/\hbar}\; Tr[U^{\dag}(\lambda,\mu) \hat{A}\hat{B}]\;d\lambda
\;d\mu \\
\end{eqnarray}
We substitute for $\hat{A}$ and $\hat{B}$ from (\ref{wc3}) and use
the fact that $A_W(q,p) = A(q,p)$ , $B_W(q,p) = B(q,p)$ to write
\begin{widetext}
\begin{eqnarray}
\nonumber C(q,p) = \frac{1}{(2\pi \hbar)^4}\int  A(q',p')
e^{i\frac{\lambda'(q-q')+\mu'(p-p')}{\hbar}}
e^{\frac{-i}{2\hbar}\left(\lambda'\mu'' - \lambda''\mu'\right)}
e^{i\frac{\lambda''(q-q'')+\mu'(p-p'')}{\hbar}} B(q'',p'')\;\; dq'
dp' dq^{\prime\prime} dp^{\prime\prime}  d\lambda^{\prime}
d\lambda ^{\prime\prime} d\mu ^{\prime} d\mu ^{\prime\prime}\\
\end{eqnarray}
\end{widetext}
where we have made use of (\ref{uu}) and
\begin{eqnarray} U^{\dag}(\lambda,\mu)
U(\lambda',\mu') = U^\dag(\lambda - \lambda',\mu - \mu')
e^{i(\lambda \mu' - \lambda' \mu)/2\hbar}.
\end{eqnarray}
The above relationship can be written as:
\begin{eqnarray}
\nonumber C(q,p) &=& A(q,p)* B(q,p)\\
&=& A(q,p)
e^{i\frac{\hbar}{2}(\frac{\overleftarrow{\partial}}{\partial
q}\frac{\overrightarrow{\partial}}{\partial p} -
\frac{\overleftarrow{\partial}}{\partial p}
\frac{\overrightarrow{\partial}}{\partial q})}B(q,p)
\end{eqnarray}
This procedure for combining two functions defines the phase space
$\ast$-product. Another way of writing it is
\begin{eqnarray}
\nonumber &\!& A(q,p)* B(q,p) =
\\ &\!& e^{\frac{i\hbar}{2}(\frac{\partial}{\partial q}
\frac{\partial}{\partial p'} - \frac{\partial}{\partial p}
\frac{\partial}{\partial q'})}A(q,p)B(q',p')
\big{|}_{(q',p')\rightarrow(q,p)}
\end{eqnarray}
Using these three entities, namely, Wigner function, Weyl
transformation and the star product, we can construct the
Wigner-Moyal-Weyl-Groenwood formalism. This formalism has been
well developed long before the recent activities in NC geometry
and been used widely for the study of semiclassical physics (see,
e.g., \cite{Liboff,Reichl,Gutzwiller}). The state of a quantum
system can be represented by a real valued function of the
canonical coordinates, the Wigner function. We notice that the
star-squared of a Wigner function (for a pure state) is
proportional to itself.
\begin{eqnarray}
W * W = \frac{1}{\hbar}W
\end{eqnarray}
The Weyl transformation of the Dirac bracket of two operators can
be shown to be equal to their commutator with respect to the star
product:
\begin{eqnarray}
[\hat A, \hat B]_W = [A , B]_* \equiv A * B - B * A
\label{starbracket}
\end{eqnarray}
 It can be shown that using the Weyl transformation of the eigenvalue
equation for the density operator, corresponding to an energy
eigen state, we obtain:
\begin{eqnarray}
H * W_n = E_n W_n
\end{eqnarray}
The eigenvalue equation is thus formulated as a ``star-gen value"
equation.

The time evolution of the system's state is governed by the
Wigner-Moyal equation. For a Hamiltonian of the form
\begin{eqnarray*}
H(x,p) = \frac{p^2}{2m} + V(x)
\end{eqnarray*}
The Wigner-Moyal equation is written as
\begin{eqnarray}
\frac{\partial W}{\partial t} &=& -\frac{2}{\hbar} \;\; W \sin
\frac{\hbar}{2}\left( \frac{\overleftarrow{\partial}}{\partial
q}\frac{\overrightarrow{\partial}}{\partial p} -
\frac{\overleftarrow{\partial}}{\partial
p}\frac{\overrightarrow{\partial }}{\partial q}\right) H\\
 &=&  \frac{2}{\hbar} \left( H \ast W-  W \ast H \right)
\end{eqnarray}
Or, equivalently, from (Eq. \ref{starbracket}),
\begin{eqnarray}
\frac{\hbar}{2} \frac{\partial W}{\partial t} = [H , W]_\ast
\end{eqnarray}

So far we have discussed everything in one space dimension. The
extension to $N$ dimensional space is straightforward. The
commutation relation takes the form:
\begin{eqnarray}
[q^i , p^j] &=& i \hbar \delta^{ij}
\end{eqnarray}
The star product is associative, that is
\begin{eqnarray}
\left[ \left(f * g \right) * h\right] &=& \left[f * \left( g * h
\right)\right],
\end{eqnarray}
The complex conjugate (c) of the star product of two functions is
given by
\begin{eqnarray}
\left( f*g \right)^c &=& g^c * f^c
\end{eqnarray}
Finally the star product of functions under integration exhibits
the cyclic property:
\begin{eqnarray}
\nonumber &\int& \left( f_1 * f_2 * \cdots * f_n \right) (x) d^N q
d^N p =
\\
&\int& \left( f_n * f_1 * \cdots * f_{n-1} \right) d^N q d^N p
\end{eqnarray}
In particular, for two functions in a 2N-dimensional phase space
(N dimensional configuration space), we have
\begin{eqnarray}
 \nonumber \int (f * g)(x)d^N q d^N p &=& \int (g * f)(x) d^N q d^N p \\
 &=& \int (f \cdot g)(x) d^N q d^N p
\end{eqnarray}
The last equation states that for two functions of phase space
coordinates ( where $[\hat{q}^i,\hat{p}_j] = i\hbar \delta^i_j$ ),
the integral of the star-product over all phase space gives the
same result as that obtained by using the ordinary product. (For
an introduction to the properties of time-independent Wigner
functions see ~\cite{curt}. Our notation in this section follows
~\cite{Hillery:1983ms}).

\section{Noncommutative Geometry}

Noncommutative geometry (NCG) has appeared in the literature ever
since Heisenberg and Snyder studied it with the hope of resolving
the ultraviolet infinity problem~\cite{snyder}. Later on it was
applied to the Landau model of electrons in a magnetic field,
where considering certain limits (the lowest energy levels) the
space of the coordinates becomes a noncommutative space. Recent
interest in noncommutative physics, however, stems from the
discovery of NCG in the context of string theory and M
theory~\cite{witten,seiberg}. NCG has been considered as a
candidate for Plank scale geometry. Hence, a successful theory of
quantum gravity may reveal the necessity or desirability of some
form of noncommutative geometry.

There are various approaches to formulate noncommutative geometry.
Early attempts using a more mathematical approach were proposed by
Alain Connes and John Madore \cite{connes1,madore2}. In this
phase, a differential NC geometry was developed and the concept of
distance and differential forms were defined. Later progress
focused more around the Wigner-Moyal formalism, described in the
last section. Almost all current work on the subject of fields in
noncommutative spaces relies on using star product and its
properties. This is the approach pursued here.

\subsection{Noncommutative $\star$-star product}

To introduce non-commutativity, one replaces the normal product
between two functions with the $\star$-product defined as
\begin{eqnarray}
f(x) \star g(x) = f(x)\;
e^{\frac{i\theta^{ij}}{2}\frac{\overleftarrow{\partial}}{\partial
x^i}\frac{\overrightarrow{\partial}}{\partial x^j}}\; g(x)
\end{eqnarray}
The $\star$-product inherits all the properties of its phase
space counterpart, the $\ast$-product.

In the previous section we used $q$ as the canonical variable for
position. From now on we denote it by $x$. In what follows, we
also use the beginning letters of the Latin alphabet, $a , b$ to
denote the coordinate indices rather than the middle letters $i
,j $. With this  we can expand the $\star$-product as
\begin{widetext}
\begin{eqnarray}
f(x) &\star& g(x) = f(x)g(x) +
\sum\limits_{n=1}\left(\frac{1}{n!}\right)
\left(\frac{i}{2}\right)^n \theta^{a_1 b_1} \ldots \theta^{a_n
b_n} \partial_{a_1} \ldots \partial_{a_k}f(x) \partial_{b_1}\ldots
\partial_{b_k}g(x)
\end{eqnarray}
\end{widetext}
In the interest of brevity, sometimes we put the lower limit of
the sum as $n=0$ to replace the first term in the above, i.e., the
$n=0$ term is  $f(x)g(x)$, without derivatives or $\theta$
dependence.

\subsection{Noncommutative Quantum Mechanics}

Using the non-relativistic limit of noncommutative quantum field
theory (NCQFT) (see~\cite{nekrasov,Szabo} for a review), one can
obtain  the Schr\"odinger equation for noncommutative quantum
mechanics (NCQM) as follows~\cite{Ho,CST}:
\begin{eqnarray}
i \hbar \frac{\partial}{\partial t} \psi = - \frac{\nabla^2}{2 m}
\psi + V \star \psi  \label{ncsch1}.
\end{eqnarray}
Here we are studying the quantum mechanics of a particle in an
external potential. As is well-known, using the form of
$\star-$product, one can write the noncommutative part as
$V\left(\hat{x}^i -
\hat{p}_j\theta^{ij}/\left(2\hbar\right)\right)\psi(x)$. However
one must pay attention to the ordering issues that can arise. To
be consistent with the definition of a $\star$-product, the
ordering here is such that all momentum operators stand to the
right of the rest of the potential and operate directly on the
wave function.

The definition of Wigner function does not change in the NC
settings. However we expect the time evolution of the Wigner
function following the Wigner-Moyal (WM) equation to be different.
To obtain the WM equation for NCQM, one can start from
(\ref{ncsch1}) in a somewhat cumbersome yet straightforward
manner. An easier way is to apply the Weyl correspondence to the
von Neumann equation,
\begin{eqnarray}
i\hbar\frac{d \hat{\rho}}{dt} = \hat{\rho} \hat{H}^{NC} -
\hat{H}^{NC}\hat{\rho}
\end{eqnarray}
where the NC Hamiltonian is written as:
\begin{widetext}
\begin{eqnarray}
\hat{H}^{NC} = \frac{\hat{p}^2}{2m} +
\sum_{n=0}\frac{1}{n!}\label{nch}
\left(\frac{-1}{2\hbar}\right)^{n} \theta^{a_1 b_1}\cdots
\theta^{a_n b_n} \partial_{a_1} \cdots
\partial_{a_n}V \,\hat{p}_{b_1} \cdots \hat{p}_{b_n}
\end{eqnarray}
\end{widetext}
We begin with the equation governing the Wigner function as it is
normally defined
\begin{eqnarray}
-i\hbar \frac{d W}{dt} = W \ast H^{NC}_W - H^{NC}_W \ast W
\end{eqnarray}
where $H^{NC}_W$ is the Weyl correspondent of the noncommutative
Hamiltonian. To find the Weyl transformation we use the usual
definition:
\begin{eqnarray}
H^{NC}_W(x,p) = 8 \int e^{2i\mathbf{y}\cdot \mathbf{p}/\hbar}
\langle x-y | \hat{H} | x+y \rangle d^3y
\end{eqnarray}
For convenience let us define
\begin{eqnarray}
A(x)^{b_1 \ldots b_n } &=& \\ \nonumber &\left(
-\frac{1}{2\hbar}\right)^n& \theta^{a_1 b_1}\cdots \theta^{a_n
b_n}\partial_{a_1} \cdots
\partial_{a_n}V(x)
\end{eqnarray}
Then we have
\begin{eqnarray}
H^{NC}_W(x,p) &=& \frac{p^2}{2m} + 8 \sum\limits_{n=0}\frac{1}{n!} \\
\nonumber &\int& e^{2i\mathbf{y}\cdot
(\mathbf{p}-\mathbf{p'}/\hbar}A(x-y)^{b_1 \ldots b_n
}p'_{b_1}\cdots p'_{b_n} d^3p'd^3y
\end{eqnarray}
which can be shown to be equivalent to
\begin{eqnarray}
\nonumber H^{NC}_W(x,p) = \frac{p^2}{2m} + 8
\sum\limits_{n=0}\frac{1}{n!} \prod\limits_{k=1}^n \left(p_{b_k} -
\frac{\hbar}{i}\partial_{b_k}\right)A(x)^{b_1 \ldots b_n }\\
\end{eqnarray}
But since $\theta^{ab}\partial_a \partial_b V$ vanishes, after
expanding the product no derivative survives and we get
\begin{eqnarray}
H^{NC}_W(x,p) = H_W(x^a-\frac{\theta^{ab}}{2\hbar}p_b,p)
\end{eqnarray}



That is, the Weyl transformation of $H(\hat{x},\hat{p})$ has the
same functional form in terms of $x$ and $p$ as the commutative
Hamiltonian but with position $x^a$ shifted by an amount
$-\frac{\theta^{ab}}{2\hbar}p_b$, where $p_b$ is the phase space
momentum.

\subsection{A Superstar $\bigstar$ Wigner-Moyal equation}

With the change of coordinates,
\begin{eqnarray}
x^{\prime a} &=& x^a - \frac{\theta^{ab}}{2\hbar}p_b \\
p^{\prime a} &=& p^a
\end{eqnarray}
we can rewrite the above equation in a more suggestive form as:
\begin{eqnarray}
 \nonumber -i\hbar\frac{\partial \tilde{W}}{\partial t} =&\,&
\tilde{W}(x',p') \bigstar H_W(x',p')\\&-& H_W(x',p') \bigstar
\tilde{W}(x',p') \,\label{ncwm}
\end{eqnarray}
with
\begin{eqnarray}
 \tilde{W}(x',p') &=& W(x^{\prime a}
+ \frac{\theta^{ab}}{2\hbar}p'_b,
p^{\prime a}) \\
\bigstar &\equiv&
e^{\frac{i\hbar}{2}\left(\overleftarrow{\partial}_\mathbf{x'}
\cdot \overrightarrow{\partial}_\mathbf{p'} -
\overleftarrow{\partial}_\mathbf{p'} \cdot
\overrightarrow{\partial}_\mathbf{x'}\right) +
\frac{i\theta^{ab}}{2}\frac{\overleftarrow{\partial}}{\partial
x^{\prime a}} \frac{\overrightarrow{\partial}}{\partial x^{\prime
b}}}
\end{eqnarray}
where $x$, $x'$,$p$ and $p'$ are phase space variables, not
operators. This is the main mathematical result of this paper.

In related works, Jing et al \cite{Jing:2004md} had derived an
explicit form for the Wigner functions in NCQM and showed that it
satisfies a generalized *-genvalue equation. (We thank Dr. J.
Prata for bringing to our attention this reference.) Dayi and
Kelleyane \cite{Dayi:2002pv} derived the Wigner functions for the
Landau problem when the plane is noncommutative. They  introduced
a generalized *-genvalue equation for this problem and found
solutions for it.

\indent Now we use this equation to examine the classical and
commutative limits. In the limit of small $\hbar$ the
equation(\ref{ncwm}) becomes
\begin{eqnarray}
\nonumber \frac{\partial \tilde{W}}{\partial t} =
&\frac{1}{2}&\left(\partial_{x^{\prime i}}H \star
\partial_{p^{\prime i}}\tilde{W} -
\partial_{p^{\prime i}}H \star \partial_{x^{\prime i}}\tilde{W} \right) -
\\\nonumber &\frac{1}{2}&\left( \partial_{x^{\prime i}}\tilde{W} \star
\partial_{p^{\prime i}}H
-\partial_{p^{\prime i}}\tilde{W} \star \partial_{x^{\prime i}}H \right)+ \\
&\frac{1}{i\hbar}& \left( H \star \tilde{W} - \tilde{W} \star H
\right)
\end{eqnarray}
{F}rom this we conclude that if $\theta$ is kept $\neq 0$ the
classical limit ($\hbar \rightarrow 0$) does not exist. In order
for the classical limit of NCQM to exist, $\theta$ must be of
order $\hbar$ or higher, or, if $\theta \rightarrow 0$ at least
as fast as $\hbar \rightarrow 0$. Note, however, that this limit
does not yield Newtonian mechanics, unless the limit of
$\theta/\hbar$ vanishes as $\theta \rightarrow 0$.

Comparing with earlier claims on this issue, a different
conclusion was reached by Acatrinei \cite{Acatrinei} who proposed
a phase-space path integral formulation of NCQM which "suggests
that a classical limit always exists" (communication from the
cited author).

Also relevant to our finding here are earlier results from
perturbative noncommutative field theories. For scalar theories,
non-planar diagrams lead to infra-red divergences \cite{BS,MRS}
which renders the theory  singular in the $\theta \rightarrow 0$
limit. This situation also arise in gauge theories (e.g.,
\cite{HMST}). In studies of perturbative NC Yang-Mills theory,
e.g., Armoni \cite{Armoni} pointed out that even at the planar
limit the $\theta \rightarrow 0$ limit of the U(N) theory does
not converge to the ordinary $SU(N) \times U(1)$ commutative
theory. This is due to the renormalization procedure being
incommensurate with noncommutativity.  This is also related to the
IR/UV issue in string theory. (There is a huge literature on NC
field theory. For reviews, see, e.g., \cite{nekrasov,Szabo})

\section{Continuity Equation and Ehrenfest Theorem}

To further explore the classical and commutative ($\hbar \approx
0$ , $\theta \approx 0$) limits, it is instructive to find out the
noncommutative version of the continuity equation and that of the
Ehrenfest theorem in such a context. Noncommutative classical
mechanics and expectation values of quantum mechanical quantities
has been studied in \cite{djami} (see earlier references therein).

\subsection{Continuity Equation}

{F}rom (\ref{ncsch1}) we have
\begin{eqnarray} \label{nc-ce1} &\,&
\frac{\partial \left( \psi^\ast \psi \right) }{\partial t}  +
\nabla \cdot \frac{\hbar}{2im}\left( \psi \nabla\psi^\ast -
\psi^\ast \nabla\psi \right) \\ \nonumber &-&
\frac{1}{i\hbar}\big( \psi^\ast \left(V\left(x\right) \star \psi
\right) - \left(\psi^\ast \star V(x) \right) \psi \big) = 0
\end{eqnarray}
To first order in $\theta$, the approximation yields
\begin{eqnarray}
\nonumber & \frac{\partial}{\partial t}&(\psi^\ast \psi) + \nabla
\cdot \\ \nonumber &\,& \left(\frac{\hbar}{2im} \left( \psi
\nabla\psi^\ast - \psi^\ast \nabla\psi \right) + \frac{1}{2\hbar}
V(x) \left( \overrightarrow{\theta} \times \nabla \left( \psi^\ast
\psi \right) \right) \right) \\ &=& 0,
\end{eqnarray}
where we have defined ($\hat{x^k}$ being the $k^{th}$ unit vector)
\begin{eqnarray}
\overrightarrow{\theta} &=& \theta_k \hat{x^k} \\
\theta_k &=& \epsilon_{ijk} \theta^{ij}
\end{eqnarray}
To this order our semi-commutative continuity equation does
suggest the following quantity as the conserved probability
current.
\begin{eqnarray}
\vec{J}_{(1)} = &\,& \nonumber \frac{\hbar}{2im}\left( \psi
\nabla\psi^\ast - \psi^\ast \nabla\psi \right) \\ &+&
\frac{1}{2\hbar} V(x) \left( \overrightarrow{\theta} \times \nabla
\left( \psi^\ast \psi \right) \right)
\end{eqnarray}
The existence of a continuity equation for all orders of $\theta$
can be inferred as follows. The $\theta$-dependent term is
proportional to
\begin{eqnarray}
\psi^\ast \left(V\left(x\right) \star \psi \right) -
\left(\psi^\ast \star V(x) \right) \psi
\end{eqnarray}
It can be shown that the difference between $A \star B$ and $AB$
is a total divergence. Using this, we can write:
\begin{eqnarray}
 \nonumber &\;& \psi^\ast \left(V\left(x\right) \star \psi \right) -
\left(\psi^\ast \star V(x) \right) \psi  \\ \nonumber &=&
\psi^\ast \star V\left(x\right) \star \psi + \partial_i Q^i -
\psi^\ast \star V\left(x\right) \star \psi - \partial_i S^i  \\
&=&
\partial_i (Q^i - S^i)
\end{eqnarray}
which shows that the $\theta$ dependent term is also a total
divergence. In fact one can explicitly compute the conserved
current to all orders. To calculate the $n^{th}$ order term, we
consider the last two terms of (\ref{nc-ce1}), where
\begin{widetext}
\begin{eqnarray}
\nonumber &&\Big[\Big( \psi^\ast \Big(V\Big(x\Big) \star \psi
\Big) - \Big(\psi^\ast \star V(x) \Big) \psi
\Big)\Big]_{n^{th}\,order} = \left(\frac{1}{n!}\right)
\left(\frac{i}{2}\right)^n \theta^{a_1 b_1} \ldots \theta^{a_n
b_n} \partial_{a_1} \\ \nonumber
 && \Big[ \Big( \psi^\ast
\partial_{a_2} \cdots
\partial_{a_n}V + (-1)^{n-1} V
\partial_{a_2} \cdots \partial_{a_n}\psi^\ast + \sum\limits_{k=2}^{n-1} \partial_{a_2} \cdots
\partial_{a_{k}}\psi^* \partial_{a_{k+1}} \cdots
\partial_{a_{n}}V\Big)\partial_{b_1} \cdots \partial_{b_n}\psi - (-1)^n
c.c. \Big] \\
\nonumber &+& V\Big(\partial_{a_1} \cdots
\partial_{a_n}\psi^\ast
\partial_{b_1} \cdots \partial_{b_n}\psi - (-1)^n
\partial_{a_1} \cdots \partial_{a_n}\psi \partial_{b_1} \cdots
\partial_{b_n}\psi^\ast \Big).
 \end{eqnarray}
 \end{widetext}
Now the two terms in the last line cancel each other, since we
can swap all $a$ indices  with $b$ indices and then bring back
the $\theta$ matrices to their original order by multiplying it
by  $(-1)^n$. The $n^{th}$ order ($n \geq 2$) result in terms of
$\theta$ for the conserved current is then given by
\begin{widetext}
\begin{eqnarray}
J^{a_1}_{(n)} = &\left( \frac{1}{i\hbar} \right)&
\left(\frac{1}{n!}\right)\left(\frac{i}{2}\right)^n \theta^{a_1
b_1} \ldots \theta^{a_n b_n} \times \\ \nonumber &\,&\Big[ \Big(
\psi^\ast
\partial_{a_2} \cdots
\partial_{a_n}V + (-1)^{n-1} V
\partial_{a_2} \cdots \partial_{a_n}\psi^\ast + \nonumber \sum\limits_{k=2}^{n-1} \partial_{a_2} \cdots
\partial_{a_{k}}\psi^* \partial_{a_{k+1}} \cdots
\partial_{a_{n}}V\Big)\partial_{b_1} \cdots \partial_{b_n}\psi - (-1)^n
c.c.\Big]
\end{eqnarray}
\end{widetext}
In the classical limit all terms must diverge unless $\theta$ is
 of order $\hbar$ or higher. One plausible argument is to assume
 that $\theta \sim \ell_p^2$, where $\ell_p$ is the Planck length ($\ell_p = \sqrt{\hbar G/c^3}$).
 In that case no term will diverge, all  terms of higher
 power in  $\theta$ will vanish and the first order term
 will be non-zero and proportional to $G/c^3$.

\subsection{Noether's theorem and Conserved Current}

Instead of performing an explicit expansion in order of $\theta$,
one can use a symmetry argument to derive a conserved current in
NCQM.  Conservation is normally linked to continuous symmetries of
the Lagrangian through Noether's theorem. One may try to trace
back both the commutative and noncommutative continuity equations
to the symmetries of a Lagrangian that produces the equations of
motion, namely the Schr\"odinger equation and its complex
conjugate. The Lagrangian for QM can be written as:

\be L = -\frac{\hbar^2}{2m} \nabla \psi^\ast \cdot \nabla \psi +
\frac{i\hbar}{2} \left(\psi^\ast \dot{\psi} - \dot{\psi}^\ast \psi
\right) + V \psi \psi^\ast. \ee This Lagrangian remains invariant
under the following transformations \be
 \delta \psi &=& i \epsilon \psi \\
 \delta \psi^\ast &=& -i \epsilon \psi^\ast
\ee The usual continuity equation in commutative QM is a
consequence of Noether's theorem.

It can be shown (for example through expanding the star product
and deriving the equations of motion, order by order) that the
following Lagrangian results in the noncommutative version of the
Schr\"odinger equation and its complex conjugate (i.e.
eq.~(\ref{ncsch1})).
\begin{eqnarray}
 L = -\frac{\hbar^2}{2m} \nabla \psi^\ast \cdot \nabla \psi &+&
\frac{i\hbar}{2} \left(\psi^\ast \dot{\psi} - \dot{\psi}^\ast \psi
\right) \\ \nonumber &+& \psi^\ast \star V \star \psi
\label{nclag}
\end{eqnarray}
One can see that this  noncommutative Lagrangian exhibits the same
symmetry as the commutative Lagrangian, and thus it admits a
conserved current to all orders of $\theta$.

\subsection{The Ehrenfest theorem}

Another way to explore the relation between quantum and classical
mechanics is the Ehrenfest theorem. What is its form in NCQM when
$\theta \neq 0$?  We begin with the time evolution of the
expectation value of $\hat{\textbf{x}}$ to lowest non-vanishing
order of $\theta$. This is given by
\begin{eqnarray}
\frac{d\langle \hat{x}^i \rangle}{dt} = \frac{\langle \hat{p}^i
\rangle}{m} + \frac{\theta^{ij}}{\hbar} \langle \partial_j V
\rangle \label{foet}
\end{eqnarray}
One can calculate this equation to all orders of $\theta$
Intuitively one can see that the form of Ehrenfest's theorem for
position follows simply from the assumption that the
noncommutative Hamiltonian can be thought of as a commutative one
in which the potential function is evaluated at a shifted
position, in a manner that was discussed above (with the
appropriate ordering). A direct calculation from first principles
confirms this view and we have:
\begin{eqnarray}
\frac{d \langle x^k\rangle}{dt} = \langle
\frac{\partial}{\hat{p}_k} \left( \frac{\hat{p}^2}{2m} +
V(\hat{x}^a-\frac{\theta^{ab}\hat{p}_b}{2\hbar})\right) \rangle
\end{eqnarray}
In fact, generally speaking, one can write the equation of motion
for the expectation value of any function of canonical operators
as:
\begin{widetext}
\be &\frac{d}{dt}& \langle f(\hat{x}^i,\hat{p}_j) \rangle =
\langle \frac{\partial f }{
\partial \hat{x}^k} \frac{\partial H(\hat{x}^a -
\frac{\theta^{ab}}{2\hbar}\hat{p}_b)}{\partial \hat{p}_k} -
\frac{\partial H(\hat{x}^a - \frac{\theta^{ab}}{2\hbar}\hat{p}_b)
}{\partial\hat{x}^k} \frac{\partial f}{\partial \hat{p}_k} \rangle
\ee
\end{widetext}
{F}rom this perspective one says that the system will behave
classically if
\begin{widetext}
\begin{eqnarray}
\label{cl}\langle \frac{\partial V(\hat{x})}{\partial \hat{x}^a}
\rangle &\approx& \frac{\partial V (\langle \hat{x}
\rangle)}{\partial \langle \hat{x}^a \rangle} \\ \langle
\partial_a
\partial_{a_1} \cdots
\partial_{a_n}V \frac{\theta^{a_1 b_1}\hat{p}_{b_1}}{2\hbar} \cdots \frac{\theta^{a_n
b_n}\hat{p}_{b_n}}{2\hbar}\rangle &\approx& \frac{\partial^n
V(\langle \hat{x} \rangle)}{\partial \langle \hat{x}^{a}
\rangle\partial \langle \hat{x}^{a_1} \rangle \cdots \partial
\langle\hat{x}^{a_n} \rangle } \left( \frac{\theta^{a_1
b_1}\langle \hat{p}_{b_1}\rangle}{2\hbar} \cdots \frac{\theta^{a_n
b_n}\langle \hat{p}_{b_n}\rangle }{2\hbar} \right)
\end{eqnarray}
\end{widetext}
These approximations improve if we consider a typical wave packet
with a spread of $\Delta x$ in position space and a spread of
$\Delta p$ in momentum space, satisfying the following conditions
as $\hbar \rightarrow 0$:
\begin{eqnarray}
\Delta x \ll \left|V \left(\frac{\partial V}{\partial x}\right)^{-1}\right| \\
\Delta p \ll \left|V\left(\frac{\partial V}{\partial
p}\right)^{-1}\right|
\end{eqnarray}
Furthermore the uncertainty principle implies that
\begin{eqnarray}
\frac{\hbar}{2} < \Delta x \Delta p \ll V^2
\left|\left(\frac{\partial V}{\partial x} \frac{\partial
V}{\partial p} \right)^{-1}\right| \label{up}
\end{eqnarray}

Going back to (\ref{foet}) we again observe that the classical
limit does not exist unless $\theta$ goes to zero at least as
fast as $\hbar \rightarrow 0$. Another important observation is
that the classical limit is NOT Newtonian mechanics, unless the
ratio $\theta/\hbar$ goes to zero as $\theta \rightarrow 0$. In
fact, assuming that $\theta \sim \ell_p^2$, the limit of $\hbar
\rightarrow 0$ gives \be \frac{d\langle \hat{x}^i \rangle}{dt}
\sim \frac{\langle \hat{p}^i \rangle}{m} +
\frac{G}{c^3}\epsilon^{ij} \langle
\partial_j V \rangle. \ee \\

{\bf Summary} In this note we have given a derivation of the
Wigner-Moyal equation under a superstar $\bigstar$ product, which
combines the phase space $\ast$ and the noncommutative $\star$ -
star products. We find that the NC-Com ($\theta \rightarrow 0$)
limit is qualitatively very different from the classical ($\hbar
\rightarrow 0$) limit. If  $\theta \neq 0$ there is no classical
limit. Classical limit exists only if $\theta \rightarrow 0$ as
least as fast as $\hbar \rightarrow 0$, but this limit does not
yield Newtonian mechanics, except when $\theta /\hbar$ vanishes in
the limit of $\theta \rightarrow 0$.

A longer paper addressing additional aspects of this issue is in preparation \cite{EHR}.\\

{\bf Acknowledgement} This work would not have been possible
without the active participation of Dr.~Albert Roura. BLH takes
pleasure to thank him for providing important insights, and AE for
his critical comments and technical help. BLH and AR would also
like to thank Dr.~Pei-Ming Ho for a useful correspondence on some
technical point in NCQM. This work is supported in part by NSF
grant PHY03-00710.



\begin{thebibliography}{99}

\bibitem{starprod} E.~P.~Wigner, ``Quantum Corrections for Thermodynamic
Equilibrium'',Phys.\ Rev.\ {\bf 40}, 749 (1932). H.~J.~Groenewold,
``On the Principles of Elementary Quantum Mechanics'', Physica
{\bf 12}, 405 (1946).  J.~E.~Moyal, ``Quantum Mechanics as a
Statistical Theory'', Proc.\ Cambridge Phil.\ Soc.\ {\bf 45} 99
(1949).

\bibitem{Drexel}
B.~L.~Hu, ``Quantum and Thermal Fluctuations, Uncertainty
Principle, Decoherence and Classicality'' Invited Talk at the
Third International Workshop on Quantum Nonintegrability, Drexel
University, Philadelphia, May, 1992.  Published in {\sl Quantum
Dynamics of Chaotic Systems,} edited by J.~M.~Yuan,  D.~H.~Feng
and G.~M.~Zaslavsky (Gordon and Breach, Philadelphia,
1993)\linebreak
\href{http://xxx.lanl.gov/abs/gr-qc/9302029}{[arXiv:
gr-qc/9302029]}.

\bibitem{HabLaf}
S.~Habib and R.~Laflamme, Phys.\ Rev.\ {\bf 42}, 4056 (1990).

\bibitem{HuZha}
B.~L.~Hu and Y.~Zhang, Mod.\ Phys.\ Lett.\ A{\bf8}, 3575 (1993);
 Int.\ J.\ Mod.\ Phys.\ {\bf 10}, 4537 (1995).

\bibitem{AnaHal}
A.~Anderson and J.~J.~Halliwell, Phys.\ Rev.\ D {\bf 48}, 2753
(1993)\href{http://xxx.lanl.gov/abs/gr-qc/9304025}{[arXiv:gr-qc/9304025]}.
C.~Anastopoulos and J.~J.~Halliwell,
   Phys.\ Rev.\ D {\bf 51}, 6870 (1995) \newline
   \href{http://xxx.lanl.gov/abs/gr-qc/9407039}{[arXiv:gr-qc/9407039]}.
   C.~Anastopoulos,
  Phys.\ Rev.\ E {\bf 53}, 4711 (1996)
  \href{http://xxx.lanl.gov/abs/quant-ph/9506031}{[arXiv:quant-ph/9506031]}.

\bibitem{ZPH} W. H. Zurek, J. P. Paz and S. Habib, Phys. Rev. Lett. {\bf 70},
1187 (1993)

\bibitem{CDS} A.~Connes, M.~R.~Douglas and A.~Schwarz, ``Noncommutative geometry and matrix theory: Compactification on tori,'', JHEP {\bf 9802}, 003 (1998)
\href{http://xxx.lanl.gov/abs/hep-th/9711162}{[arXiv:hep-th/9711162]}.

\bibitem{witten} E.~Witten,``Noncommutative Geometry And String Field Theory,''
  Nucl.\ Phys.\ B {\bf 268}, 253 (1986).

\bibitem{seiberg} N.~Seiberg and E.~Witten, ``String Theory and Noncommutative
Geometry'', JHEP {\bf 9909} (1999) 032
\href{http://xxx.lanl.gov/abs/hep-th/9908142}{[arXiv:hep-th/9908142]}.

\bibitem{konechny}A.~Konechny and A.~Schwarz,
 ``Introduction to M(atrix) theory and noncommutative geometry. Parts I and II,''
  Phys.\ Rept.\  {\bf 360}, 353 (2002)
  \href{http://xxx.lanl.gov/abs/hep-th/0012145}{[arXiv:hep-th/0012145]}, \newline
  \href{http://xxx.lanl.gov/abs/hep-th/0107251}{[arXiv:hep-th/0107251]}.
\bibitem{douglas} M.~R.~Douglas, ``Two Lectures on D-Geometry and
Noncommutative Geometry'', in: {\sl Nonperturbative Aspects of
String Theory and Supersymmetric Gauge Theories}, eds. M.~J.~Duff,
E.~Sezgin, C.~N.~Pope, B.~R.~Greene, J.~Louis, K.~S.~Narain,
S.~Randjbar-Daemi and G.~Thompson (World Scientific, Singapore,
1999), p. 131
\href{http://xxx.lanl.gov/abs/hep-th/9901146}{[arXiv:hep-th/9901146]}.

\bibitem{nekrasov} Michael R.~Douglas and Nikita A.~Nekrasov, ``Noncommutative Field
Theory''
 \href{http://xxx.lanl.gov/abs/hep-th/0106048}{[arXiv:hep-th/0106048]}.

\bibitem{Majid} S.~Majid, ``Hopf Algebras for Physics at the Planck
Scale", Class.\ Quant.\ Grav.\ {\bf 5} (1988) 1587. S. Majid,
 {\sl Foundations of Quantum Group Theory}, (Cambridge University Press,
Cambridge 1995).

\bibitem{Liboff} R.~Liboff, {\sl Kinetic Theory}, (John Wiley, New York,
1998).

\bibitem{Reichl} L.~E.~Reichl, {\sl The Transition to Chaos: Conservative Classical
Systems and Quantum Manifestations} (Springer-Verlag, 2004).

\bibitem{Gutzwiller} Martin C.~Gutzwiller, {\sl Chaos in Classical and Quantum
Mechanics} (Interdisciplinary Applied Mathematics)
(Springer-Verlag, Berllin, 1991).

\bibitem{Hillery:1983ms}  M.~Hillery, R.~F.~O'Connell, M.~O.~Scully and
E.~P.~Wigner,``Distribution Functions In Physics: Fundamentals,''
Phys.\ Rept.\  {\bf 106}, 121 (1984).

\bibitem{weyl} H.~Weyl, The Theory of Groups and Quantum Mechanics, Dover, New York
(1950).

\bibitem{harvey} J.~A.~Harvey,``Komaba lectures on noncommutative solitons and D-branes,''
\href{http://xxx.lanl.gov/abs/hep-th/0102076}{[arXiv:hep-th/0102076]}.

\bibitem{curt}T.~Curtright, D.~Fairlie and C.~K.~Zachos,
Phys.\ Rev.\ D {\bf 58}, 025002 (1998)
\href{http://xxx.lanl.gov/abs/hep-th/9711183}{[arXiv:hep-th/9711183]}.

\bibitem{snyder} H.~Snyder, Phys.\ Rev.\ {\bf 71}, 38 (1947).

\bibitem{connes1} A.~Connes, {\sl Noncommutative Geometry}  (Academic, New York,
1994).

\bibitem{madore2} J.~Madore, {\sl An introduction to noncommutative differential geometry and its physical
applications} No. 257 in London Mathematical Society Lecture Note
Series. 2nd edn. (Cambridge University, 1999).

\bibitem{CST} M.~Chaichian, M.~M.~Sheikh-Jabbari, and A.~Tureanu. \href{http://link.aps.org/abstract/PRL/v86/p2716}
{Phys.\ Rev.\ Lett.\ {\bf 86} 2716}.

\bibitem{Szabo} R.~J.~Szabo,
``Quantum Field Theory on Noncommutative Spaces," Phys. Rep. {\bf
378}, 207-299  (2003).

\bibitem{Ho} P.~M.~Ho and H.~C.~Kao,
  ``Noncommutative quantum mechanics from noncommutative quantum field
  theory,'' \newline \href{http://link.aps.org/abstract/PRL/v88/e151602}{Phys.\ Rev.\ Lett.\  {\bf 88}, 151602
  (2002)}\newline\href{http://xxx.lanl.gov/abs/hep-th/0110191}{[arXiv:hep-th/0110191]}.
\bibitem{nccm}G.~D.~Barbosa,``On the meaning of the string inspired noncommutativity and its
  implications,'' \href{http://jhep.sissa.it/archive/papers/jhep052003024/jhep052003024.pdf}{JHEP {\bf 0305}, 024
  (2003)}.
\bibitem{Jing:2004md}
  S.~C.~Jing, F.~Zuo and T.~H.~Heng,
  JHEP {\bf 0410}, 049 (2004).
\bibitem{Dayi:2002pv}
  O.~F.~Dayi and L.~T.~Kelleyane,
  Mod.\ Phys.\ Lett.\ A {\bf 17}, 1937 (2002)
  [arXiv:hep-th/0202062].

\bibitem{djami}A.E.F. Djemai, ``On Noncommutative Classical Mechanics",
\href{http://xxx.lanl.gov/abs/hep-th/0309034}{[arXiv:hep-th/0309034]}.

\bibitem{Acatrinei}
 C.~Acatrinei, JHEP {\bf 0109} (2001) 007

\bibitem{BS} D. Bigatti and L. Susskind, 
\href{http://xxx.lanl.gov/abs/hep-th/9908056}{[arXiv:hep-th/9908056]}

\bibitem{MRS} S. Minwalla, M. Van Raamsdonk and N. Seiberg, {''Noncommutative Perturbative Dynamics''},
\href{http://xxx.lanl.gov/abs/hep-th/9912072}{[arXiv:hep-th/9912072]}.
M. V. Raamsdonk and N. Seiberg, 
JHEP {\bf 0003} (2000) 035.

\bibitem{HMST} e.g., M. Hayakawa, 
Phys.\ Lett.\ B {\bf 478} (2000) 394;
\href{http://xxx.lanl.gov/abs/hep-th/9912167}{[arXiv:hep-th/9912167]}.
 A. Matusis, L. Susskind and N. Toumbas, 
\href{http://xxx.lanl.gov/abs/hep-th/0002075}{[arXiv:hep-th/0002075]}.

\bibitem{Armoni} A. Armoni, Nucl.\ Phys.\ B {\bf 593} (2001) 229-242.

\bibitem{EHR}Ardeshir Eftekharzadeh, B.~L.~Hu and
Albert Roura, ``Noncommutative Geometry and Quantum-Classical
Correpondence" (in preparation).

\end{thebibliography}
\end{document}